\documentclass[%
 reprint,
nofootinbib,
 amsmath,amssymb,
 aps,
 prd,
 longbibliography
]{revtex4-1}
\usepackage[colorlinks=true, citecolor=blue, linkcolor = black]{hyperref}
\usepackage{graphicx}

\usepackage[obeyFinal]{easy-todo}

\newcommand{\bra}[1]{\langle #1 |}
\newcommand{\ket}[1]{| #1 \rangle}
\newcommand{\cM}{\mathcal{M}}
\newcommand{\cR}{\mathcal{R}}

\renewcommand{\Re}{\operatorname{Re}}
\renewcommand{\Im}{\operatorname{Im}}

\begin{document}

\preprint{INR-TH}

\title{Uncertainties of hadronic scalar decay calculations}
\author{F. Bezrukov}
\email{Fedor.Bezrukov@manchester.ac.uk}
\affiliation{The University of Manchester, School of Physics and
  Astronomy, Oxford Road, Manchester M13 9PL, United Kingdom}

\author{D. Gorbunov}
\email{gorby@ms2.inr.ac.ru}
\affiliation{Institute for Nuclear Research of the Russian Academy of
  Sciences, Moscow 117312, Russia}%
\affiliation{Moscow Institute of Physics and Technology, Dolgoprudny
  141700, Russia}%

\author{I. Timiryasov}
\email{inar.timiryasov@epfl.ch}
\affiliation{Institute of Physics, Laboratory for Particle Physics and Cosmology,\\
  \'{E}cole Polytechnique F\'{e}d\'{e}rale de Lausanne, CH-1015
  Lausanne, Switzerland}

\begin{abstract}
  Dispersion relations---a corollary of unitarity and
  analyticity---are among the few methods which can be employed to
  study low energy processes in QCD. In this work we critically
  revisit the calculation of decay widths of a hypothetical light
  scalar boson into pions and kaons. Strong interactions of the mesons
  in the final state affect these decays significantly.  We argue that
  applications of the dispersion relations to the calculation of these
  decay widths, which are present in the literature, rely on an
  uncontrollable approximation of the reduced S-matrix. When a limited
  number of initial and final decay channels are considered, in general neither a
  unitary S-matrix nor its version corrected by an inelasticity factor
  are justified. The scalar form factor calculation
  provides a transparent example to illustrate this statement.
  Therefore, the results of this calculation, which are important for
  theoretical, cosmological and experimental applications, must be
  treated with caution. 
  Our work, therefore, calls for  a breakthrough
  in the understanding of the dispersion relations of realistic
  many-channel systems.
\end{abstract}

\maketitle

\section{Introduction\label{sec:intro}}

Light scalar bosons, singlet with respect to the gauge group of the
Standard Model of particle physics (SM), appear in many
phenomenological models, see
e.g.~\cite{Gorbunov:2000th,Foot:2011et,Bezrukov:2013fca}.  They became
a part of physics programs of current and planned experiments (see,
e.g.\ the SHiP \cite{Alekhin:2015byh}, MATHUSLA \cite{Curtin:2018mvb}
and FASER \cite{Ariga:2018uku} physics papers).  From the perspectives
of direct experimental search, knowing the decay widths and branching
ratios of such bosons is crucial since they determine the lifetimes
(decay length) and detection signatures respectively.

The decay width (and branching ratios consequently) of the light
scalar boson with mass around $1$\,GeV suffers from the uncertainties
originating from a lack of understanding of its hadronic decays
\cite{Bezrukov:2009yw,Astapov:2015otc,Choi:2016luu,Fradette:2017sdd}.
One of the methods widely applied with the aim of reducing these
uncertainties is the dispersive approach.  In particular it was
applied to study the hadronic decay channels of light scalar bosons in
relation to the phenomenology of a Higgs boson $H$, see e.g.\
\cite{Ellis:1975ap,Shifman:1978zn,Voloshin:1980zf,
  Voloshin:1985tc,Raby:1988qf,Truong:1989my,Donoghue:1990xh}.  Even
though there was some controversy in the literature, see, e.g.\
discussion in~\cite{Clarke:2013aya}, the method of
\cite{Donoghue:1990xh} seems to be the most advanced and accurate and
has been widely adopted to estimate the hadronic decay rates of
GeV-scale exotic scalars
\cite{Bezrukov:2009yw,Winkler:2018qyg,Monin:2018lee}.  However, the
question of reliability of the dispersive approach is not studied in
detail.

The reason to question the reliability of the approach is the
following.  The dispersion relations, based on analyticity, which is
in turn a consequence of causality, allow for reconstructing an
amplitude $A$ from its imaginary part,
\begin{equation}
  A(s) = \frac{1}{\pi}\int_0^\infty ds' \frac{\Im A(s')}{s' - s - i \epsilon}.
  \label{disp1}
\end{equation}
The amplitude of the decay of the scalar boson $S\to X \bar{X}$,
$X = \pi, K, \ldots$ is proportional to a linear combination of form
factors $F(s)=\bra{ X(q) \bar{X}(q')} j \ket{0}, \; s=(q+q')^2$, where
$j$ is the scalar current, i.e.\ $j= \sum_{\mathrm{quarks}} \bar{q}q$.
The Watson theorem \cite{Watson:1954uc}, which is based on unitarity,
states that the phase of the form factor $\alpha$ equals the phase
shift $\delta$ of the scattering amplitude $X\bar X\to X\bar X$.  This
observation transforms \eqref{disp1} into an integral equation
determining the form factor $F$ for a given function $\delta$,
\begin{equation}
  \Re F(s) = \frac{1}{\pi} \mathrm{P}\!\!\int_0^\infty ds' \frac{\Re F(s')
    \tan(\delta)}{s' - s }, 
  \label{integral_equation}
\end{equation}
where symbol $\mathrm{P}$ indicates the principal value integral and 
the real and imaginary parts of $F$ are related by the phase $\alpha$.
In order to determine the form factor $F$ one needs to know the function
$\delta(s)$ up to infinite energies. Meanwhile, in practice, the phase
shifts $\delta$ are measured only in the limited energy intervals.

How sensitive is the approach to the behaviour of $\delta$ at large
energies?  Naively one can expect that the effect of small variations
of the phase $\delta$ at large $s'$ is suppressed by the factor
$1/(s' - s)$.  However, the variations of the phase are not
necessarily small. For instance, if a resonance is present at some
energy $\sqrt{s_r}$, the phase shift $\delta$ increases by $\pi$
around $s_r$. In this case the solution of
Eq.~\eqref{integral_equation} exhibits a pronounced peak.

This is, however, not the only problem. Indeed, the equation
\eqref{integral_equation} was obtained under the assumption of
validity of the Watson theorem. However, the theorem is formulated for
a single reaction channel (e.g.\ $\pi\pi\to\pi\pi$).  In the presence
of inelastic channels (such as $\pi\pi\to KK$) it should be modified
to include these extra channels. The single equation
\eqref{integral_equation} becomes a system of coupled singular
integral equations. Most of the analyses up to date were restricted to
  the two channel case. So another question is how the inelastic
channels, which are not accounted for in the two channel
consideration, modify the result.  We argue here that the answer to
this question is non-trivial, and that the extra channels cannot be
neglected in general case.

This work is organized as follows.
Section~\ref{sec:review_of_the_formalism} contains an overview of the
calculation of the scalar decay rates into hadrons.  We define the
main objects of interests, namely the form factors in
Section~\ref{sub:scalar_form_factors}. The dispersion relations and
unitarity are discussed in
Section~\ref{sub:dispersion_relations_and_unitarity}.  In
Section~\ref{sec:reduction_in_the_number_of_channels} we describe the
reduction of an $\cM$ channel system to an $\cR$ channel system with
$\cR < \cM$. We argue that the \emph{unitary} $\cR$ channel system
cannot properly describe the behaviour of the form factors, which can
only be calculated in the full $\cM$ channel system.  We illustrate
this statement on the example of $\cM=2 \to \cR=1$ reduction. This
demonstration implies that within any treatements suggested so far
the uncertainties in the hadronic channels remain untamed and may
reach a factor of order ten, contrary to the recent 
report~\cite{Winkler:2018qyg}.
We discuss the implications of our
statement in Section \ref{sec:Conclusion}.


\section{Review of the formalism} 
\label{sec:review_of_the_formalism}

In this Section we define the Lagrangian of the model and introduce
the form factors following the notations of
Ref.~\cite{Donoghue:1990xh}.  At low energies, where the Chiral
Perturbative Theory (ChPT) is reliable, it can be used to calculate
these form factors.  At higher energies the dispersive approach is
employed.  We present a brief review of the dispersion methods and
sketch the analysis of Ref.~\cite{Donoghue:1990xh}.

\subsection{Scalar form factors} 
\label{sub:scalar_form_factors}

Imagine a light scalar $S$ mixed with the SM Higgs scalar.  If the
mass of $S$ is well below the electroweak scale the Higgs boson can be
integrated out.  The renormalizable low energy interaction Lagrangian
then reads
\begin{equation}
  \mathcal{L}_{\mathrm{int}}  = - \sum_f\frac{g_* m_f}{v}\bar{\psi}_f \psi_f\,S,
  \label{Lagr}
\end{equation}
where the sum runs over all SM fermions $f$ with masses $m_f$,
$v\simeq 246$\,GeV, and the effective coupling $g_*$ originates from
the mixing of $S$ with the SM Higgs.

Lagrangian~\eqref{Lagr} controls decay channels of the scalar $S$.
While the leptonic decay rates can be readily calculated, this is not
the case for the hadronic channels, if the scalar is at the GeV-scale.
The reason is that final state interactions of the strongly
interacting particles alter the result significantly.  Recall, the amplitude of
the process $S\to X \bar{X}$, $X = \pi, K, \ldots$ is proportional to
a linear combination of the form factors
$F(s)=\bra{ X(q) \bar{X}(q')} j \ket{0}, \; s=(q+q')^2$, where $j$ is
the scalar current, i.e.\ $j= \sum_{\mathrm{quarks}} \bar{q}q$.  The
contribution of light quarks can be addressed directly, whereas heavy
quarks can be accounted for using the trick by Shifman, Vainshtein,
and Zakharov \cite{Vainshtein:1980ea}. In short, the scalar currents
of heavy quarks are related to those of light quarks via the trace of
energy-momentum tensor $\theta_\mu^\mu$.

In Ref.~\cite{Donoghue:1990xh} only the decay channels to pions and
kaons have been considered.  Below we refer to this as the two channel
case. Analogously, in the single channel case only pions are
considered.

The independent form factors responsible for the decay of $S$ in the two channel case are
\begin{subequations}
\begin{align}
\bra{\pi^i(p)\pi^k(p')} \theta_\mu^\mu \ket{0} &\equiv \Theta_\pi(s) \delta^{i k},
\label{Theta}\\
\bra{\pi^i(p)\pi^k(p')} m_u \bar{u}u + m_d \bar{d}d \ket{0} &\equiv 
\Gamma_\pi(s) \delta^{i k}, \label{Gamma}\\
\bra{\pi^i(p)\pi^k(p')} m_s \bar{s}s \ket{0} &\equiv \Delta_\pi(s) \delta^{i k},
\label{Delta}
\end{align}
\label{FF_definition}\end{subequations}
where $s = (p+p')^2$.  Kaon form factors are defined analogously.  The
decay width $S\to \pi\pi$ is proportional \cite{Donoghue:1990xh} to
the square of
\begin{equation}
  G_\pi(s) = \frac29 \Theta_\pi(s)+ \frac79 \left( \Gamma_\pi(s) + \Delta_\pi(s) \right). 
  \label{G}
\end{equation}
An analogous combination is defined for kaons.

At sufficiently low energies, the values of the form factors defined
above can be computed using the ChPT.  To the leading order in momentum
($\mathcal{L}_2$ order) one gets \cite{Donoghue:1990xh}
\begin{subequations}
\begin{align}
\Theta_\pi(s)  &= s + 2m_\pi^2, & \Theta_K(s)  &= s + 2m_K^2,
\label{Theta0}\\
\Gamma_\pi(s)  &= m_\pi^2,      & \Gamma_K(s)  &= \frac12 m_\pi^2,
\label{Gamma0}\\
\Delta_\pi(s)  &= 0,           & \Delta_K(s)  & = m_K^2 - \frac12 m_\pi^2.
\label{Delta0}
\end{align}
\label{ChPT_FF}\end{subequations}
In the next subsection we describe how to determine the form factors
\eqref{FF_definition} at larger $s$.

\subsection{Dispersion relations and unitarity} 
\label{sub:dispersion_relations_and_unitarity}

In this subsection we discuss how unitarity and analyticity can be
used to reconstruct the form factors from the scattering data.

The scattering of asymptotic states, such as pions, kaons, eta mesons,
etc., is described by an unitary S-matrix.  The standard definition
of the T-matrix is
\begin{equation}
  S_{i j}  = \delta_{i j} + 2 i \sqrt{\sigma_i \sigma_j} T_{i j},
  \label{Smatrix0}
\end{equation}
where 
\begin{equation}
  \sigma_i(s) = \sqrt{1 - 4 m_i^2/s} \,\theta(s - 4m_i^2).
  \label{sigma}
\end{equation}
The fundamental quark interactions for each channel $i$ correspond to
the form factors $F_i(s)$.  In the two channel case, with $i=1,2$
denoting pions and kaons respectively (so that they match 
with Eqs.~\eqref{FF_definition}
and \eqref{ChPT_FF}),
these are
\begin{equation}
  F_1(s)= \bra{0} \hat{X_0} \ket{\pi \bar{\pi}}, \quad
  F_2(s)= \bra{0} \hat{X_0} \ket{K \bar{K}},
  \label{formfactors}
\end{equation}
where $ \hat{X_0}$ is a scalar operator of zero isospin, e.g.\
$m_u\bar{u}u+m_d\bar{d}d$, see Eq.~\eqref{FF_definition}.  For the form factors of this type, one can
derive the unitarity conditions
\begin{equation}
  \Im F_i = \sum_{j = 1}^n T^*_{i j} \sigma_j(s) F_j(s).
  \label{unitarity}
\end{equation}
Note that the conditions \eqref{unitarity} are valid for the general
case when $n$ channels are present.

The form factor can be reconstructed from its imaginary part using
\eqref{disp1}.  This leads to a set of $n$ coupled integral equations
with singular kernels known as the Muskhelishvili--Omn{\`e}s equations
\begin{equation}
  F_i(s) = \frac{1}{\pi} \sum_{j = 1}^n \int\limits_{4 m_j^2}^\infty
   \frac{ds'}{s'-s}T^*_{i j}(s')\sigma_j(s') F_j(s'), \;\; i = 1\ldots n.
  \label{MO}
\end{equation}
One can see immediately that Eqs.~\eqref{MO} \emph{require the off-diagonal
elements of $T$ in the unphysical region} (that is below the
corresponding threshold).

The single channel case is an important basic situation.  If only one
final state is considered, the unitary scattering matrix is fully
determined by the phase shift $\delta$, $S= \exp(2i \delta(s))$, and
Eq.~\eqref{unitarity} simplifies to
\begin{equation}
  \Im F(s) = - \left( e^{-2i \delta} - 1 \right) F(s).
  \label{single_channel_unitarity}
\end{equation}
The form factor is a real analytic function of $s$ in the plane with a
cut along the real axis starting from the $4 m^2$ threshold.  For the
real values of $s$ we define $F(s) \equiv F(s+i \epsilon)$. Since
$F(s^*) = F^*(s)$, the imaginary part of $F$ is related to the
discontinuity as
$\Im F(s + i \epsilon) = 2i \left( F(s + i \epsilon) - F(s - i
  \epsilon) \right) $.  Therefore, eq~\eqref{single_channel_unitarity}
can be rewritten as
\begin{equation}
  F(s + i \epsilon) = F(s - i \epsilon)e^{2i \delta}.
  \label{manifest_watson}
\end{equation}
Thus the phase of the form factor is equal to the phase shift of
elastic scattering amplitude. 
This relation is known as the Watson theorem~\cite{Watson:1954uc}.

Noticing that the factor
$T^* \sigma = \left( \exp(2i \delta) - 1 \right)/2i $ in
Eq.~\eqref{unitarity} can be rewritten as
$\left( \exp(2i \delta) - 1 \right)/2i = \exp(i \delta) \sin \delta$,
we get the standard form of the Omn{\`e}s equation (we present here a
once subtracted form)
\begin{equation}
    F(z) =F(0) + \frac{s}{\pi}\int_{4 m^2}^\infty \frac{ds}{s(s-z)} e^{i \delta(s)} \sin\delta(s)
  F^*(s).
  \label{MO_sub}
\end{equation}
The single channel equation \eqref{MO_sub} has an analytic solution
\cite{Omnes:1958hv}
\begin{equation}
  F(z) = F(0) \exp \left \{ \frac{z}{\pi} \int_{4 m^2}^\infty ds \frac{ \delta(s)}{s(s-z)}
   \right \}.
   \label{Omn_sol}
\end{equation}
In fact, a general solution is given by
\begin{equation}
  F(z) = P(z) \Omega(z),
\end{equation}
where $P(z)$ is a polynomial in $z$ and the Omn{\`e}s function reads
\begin{equation}
  \Omega(z) = \exp \left \{ \frac{z}{\pi} \int_{4 m^2}^\infty ds \frac{ \delta(s)}{s(s-z)}
  \right \}.
  \label{Omn}
\end{equation}

The general analytic solution to the system \eqref{MO} for $n>1$ 
cannot be obtained in a closed form. However, it is known that the system
has $n$ independent (fundamental) solutions~\cite{muskhelishvili}.
Let us denote these solutions as $F_i^{(j)}$, where $i$ indicates the
final state, i.e.\ $i=\pi, K, \ldots$, whilst $j$ enumerates the
independent solutions.  One can combine $n$ fundamental solutions
$F_i^{(j)}$ into a matrix.  Let us denote the determinant of this
matrix as $\hat{F}$.  From \eqref{Smatrix0} and \eqref{unitarity} one
can find that
\begin{equation}
  \hat{F} = \det(S) \hat{F}^\ast.
  \label{eq_det}
\end{equation}
Unitarity of the S-matrix implies that $\det S = \exp(2 i \Delta)$,
where $\Delta$ is the sum of the phase shifts. Therefore the
relation~\eqref{eq_det} has precisely the form of the single channel
equation~\eqref{manifest_watson}. Solution~\eqref{Omn} allows for
performing a cross check of the numerical solution of the $n$ channel
case.  Moreover, the analytical solution allows one to determine the
behaviour of the $\hat{F}$ at infinity. It is controlled by the value
of $\Delta(\infty)$ and reads
\begin{equation}
  \hat{F}(s)\sim s^{-\Delta(\infty)/\pi}.
  \label{asympt}
\end{equation}

\subsection{Two channel case} 
\label{sub:two_channel_case}

In this Section we limit ourselves to the two channel case, that is,
only pions and kaons are considered.  For the sake of simplicity, we
omit all the Clebsch--Gordan coefficients.  The standard
parametrization of the $2 \times 2$ unitary S-matrix is
\begin{equation}
S = 
\begin{pmatrix}
  \eta\, e^{2 i \delta_\pi} & i\left(1-\eta^2\right)^{1/2} e^{i (\delta_\pi+\delta_K)}\\
  i\left(1-\eta^2\right)^{1/2} e^{i (\delta_\pi+\delta_K)} & \eta\, e^{2 i \delta_K}
\end{pmatrix},
\label{Smatrix}
\end{equation}
where the phases $\delta_\pi$, $\delta_K$ and the inelasticity $\eta$
are functions of $s$ which should be obtained from experimental data.
As we have mentioned above, the off-diagonal elements of $T(s)$ are
needed both in the physical and unphysical regions. Therefore,
analytic parametrisations allowing for extrapolation of $T_{21}$ below
$s=4m_K^2$ are needed.  There are several empiric parametrisations of
this type~\cite{Au:1986vs,Kaminski:1997gc,Kaminski:1998ns}.

As we have already mentioned, there are no known solutions for $n>1$
channels, so the system \eqref{MO} should be solved numerically.  This
is a challenging task since one has to deal with singular kernels. The
principal value integrals should be treated carefully.  There are two
different approaches to the problem, the iterative one of
Ref.~\cite{Donoghue:1990xh} and the one of
Ref.~\cite{Moussallam:1999aq}, which is based on the singular value
decomposition.  Our numerical studies is based on the latter method.

One can construct two linearly independent solutions $C_i$ and $D_i$
to the system \eqref{MO} as
\begin{equation}
  C_i(s)|_{s=0} = \delta_{i 1}, \quad D_i(s)|_{s=0} = \delta_{i 2}.
  \label{fundamental_solutions}
\end{equation}
These solutions are combined to get the form factors \eqref{Theta},
\eqref{Gamma} and \eqref{Delta}. The coefficients in the linear
combinations are fixed using the ChPT results \eqref{Theta0},
\eqref{Gamma0} and \eqref{Delta0} as
\begin{equation}
  \begin{aligned}
    \Gamma_\pi(s) &= \Gamma_\pi(0)  C_1(s) + \Gamma_K(0) D_1(s),\\ 
    \Gamma_K(s)   &= \Gamma_\pi(0)  C_2(s) + \Gamma_K(0) D_2(s),\\
  \end{aligned}
  \label{linear_combinations}
\end{equation}
and analogously for other form factors
\cite{Donoghue:1990xh}. 
In principle, Eqs.~\eqref{linear_combinations}
define the solution for the form factors at any values $s$.

Let us note in passing that it has not been rigorously demonstrated that the methods of 
Refs.~\cite{Donoghue:1990xh}
and \cite{Moussallam:1999aq} guarantee the proper behaviour of the solutions $C_i$ and $D_i$ at infinity~\cite{muskhelishvili} provided that only \eqref{fundamental_solutions} are
 required.


\section{Reducing the number of channels} 
\label{sec:reduction_in_the_number_of_channels}

In this Section we consider implications of the restriction to a certain number of channels.

\subsection{General discussion} 
\label{sub:general_discussion}

Suppose that the system can be fully described by a $\cM$ channel
S-matrix (where $\cM$ can be infinite).  Of course, exact solution for
such a system for large $\cM$ is impossible.  Therefore, in real world
applications the number of channels is limited.  Let us consider
$\cR < \cM$ channels.  In order to make the discussion somewhat more
transparent we write down the system of equations explicitly
\begin{widetext}
\begin{equation}
\begin{array}{c@{\;=\;}c}
F_1(s) & \sum\limits_{j = 1}^\cR \int \frac{ds'}{\pi} 
\frac{T^*_{1 j}(s') \sigma_j(s')}{s'-s} F_j(s')
+\left\{ \sum\limits_{j = \cR + 1}^\cM \int \frac{ds'}{\pi} 
\frac{T^*_{1 j}(s') \sigma_j(s')}{s'-s} F_j(s')\right\}\\
F_2(s) & \sum\limits_{j = 1}^\cR \int \frac{ds'}{\pi} 
\frac{T^*_{2 j}(s') \sigma_j(s')}{s'-s} F_j(s')
+\left\{\sum\limits_{j = \cR + 1}^\cM \int \frac{ds'}{\pi} 
\frac{T^*_{2 j}(s') \sigma_j(s')}{s'-s} F_j(s')\right\}\\
 \hdotsfor{2}\\
F_\cR(s) & \sum\limits_{j = 1}^\cR \int \frac{ds'}{\pi} 
\frac{T^*_{\cM j}(s') \sigma_j(s')}{s'-s} F_j(s')
+\left\{\sum\limits_{j = \cR + 1}^\cM \int \frac{ds'}{\pi} 
\frac{T^*_{\cR j}(s') \sigma_j(s')}{s'-s} F_j(s')\right\}
\end{array}
\label{Reduced_system}
\end{equation}
\end{widetext}
The terms in braces depend on the form-factors with $j>\cR$ and can
only be found from the \emph{full} system of equations.  The only
generic statement which can be done is that these undetermined terms
are suppressed as $(4m_\cR^2-s)/(4m_{\cR+j}^2-s)$ which follows from
the fact that $\sigma_j(s)|_{s<4m_j^2} = 0$, see Eq.~\ref{sigma}.  In
particular, $\eta \eta$ channel is suppressed rather weakly compared
to the KK channel, $(m_K/m_\eta)^2\simeq 0.8$.  Therefore, the
corrections originated from the terms in braces cannot be estimated
\emph{a priori.}  However, considering the $\cR$ channel system
precisely means that all the terms in braces are set equal to zero.
Validity of this assumption may be justified only far below the
thresholds of the neglected channels (or there is a particular reason
to believe that some channel has suppressed T-matrix elements, like
the phase space suppression of the $4\pi$ channel at low
energies).

Usual assumption for the reduction to $\cR$ channels is based on the
observation that below the $\cR+1$ threshold the $\cR\times\cR$
S-matrix is unitary.  However, as we see from (\ref{Reduced_system})
the contribution of the high energy channels do not disappear below
their threshold, making this reduction impossible.\footnote{There is a
  recent work \cite{Ropertz:2018stk} which offers a parametrization
  for the strange scalar form factor $\Delta_\pi$ that is claimed to be valid up to
  2\,GeV. It is tempting to check if this approach can be generalized
  to other form factors.} Below we illustrate this
statement with the simplest example.

\subsection{Example of 2 to 1 reduction} 
\label{sub:example_of_2_to_1_reduction}

As we have stressed above, there is no control of an error arising
from the reduction of  channels.  The error can be
obtained by the direct comparison of the solutions of the system with
$\cM$ and $\cR$ channels. We perform such a comparison taking
$\cM = 2$ and $\cR = 1$.

The two channel case is solved numerically using the method based on
Ref.~\cite{Moussallam:1999aq}.  We use the parametrisation of the 
T-matrix by Truong and Willey \cite{Truong:1989my}.\footnote{As was
  pointed in Ref.~\cite{Donoghue:1990xh}, the sign of the parameter
  $\lambda$ in the parametrisation must be flipped in order to fit the
  experimental data, so we use $\lambda = -0.1$.}  This
parametrisation admits an analytical solution and thus allows us to cross
check the numerical results.

\begin{figure}[!htb]
\hskip 0.05\columnwidth
\includegraphics[width=0.9\columnwidth]{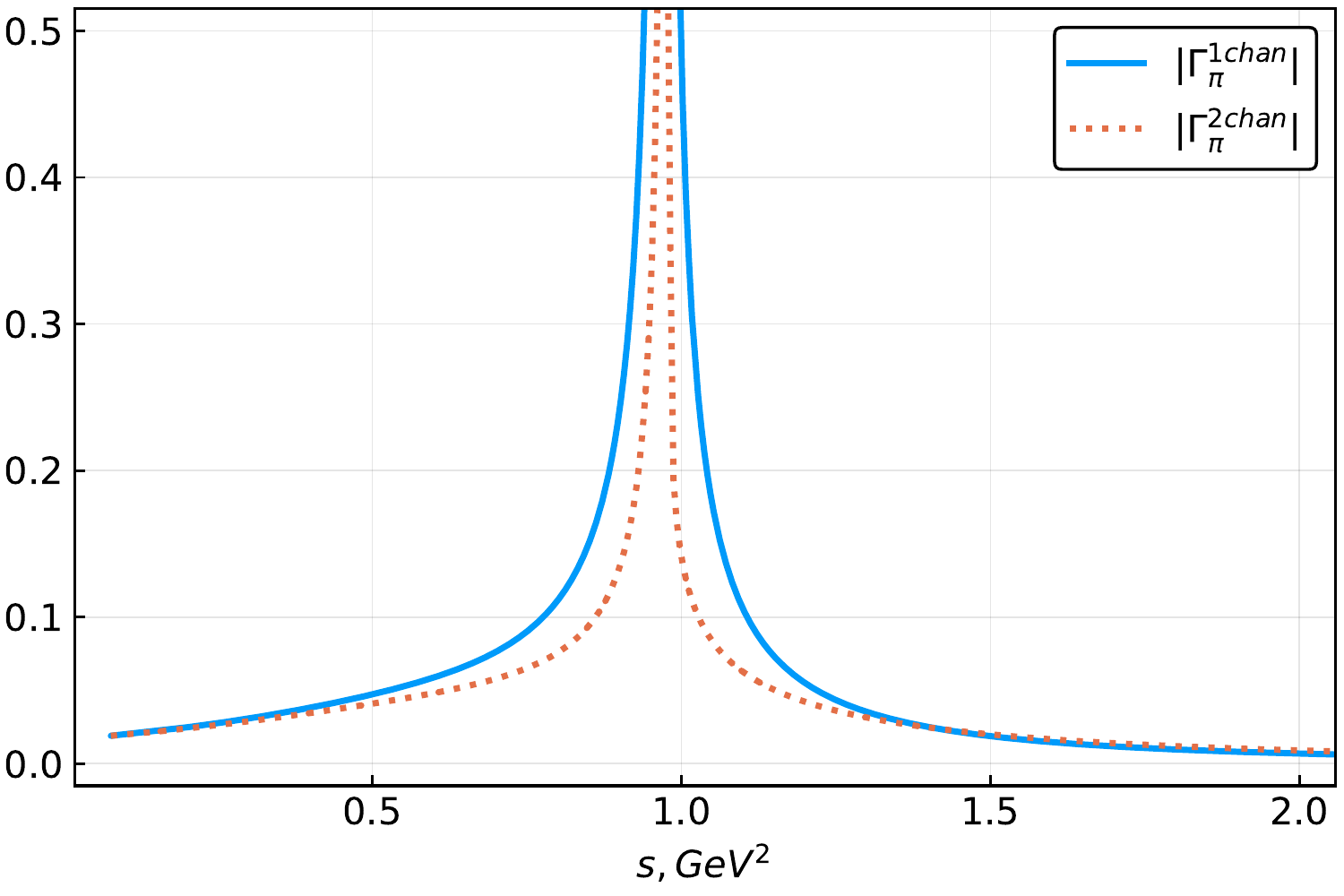}
\caption{Absolute values of the pion form factors as functions of $s$.
The solid line represents  $|\Gamma_\pi(s)|$ obtained in the single channel approximation
(Eq.~\eqref{Omn_sol}) and normalized at  $s=4m_\pi^2$ according to~\eqref{Gamma0}.
The dotted line is the analogous form factor in the two channel approximation.
\label{fig:FF}}
\end{figure}

In the single channel case, the solution is given by
Eq.~\eqref{Omn_sol}, where the phase $\delta(s)$ is obtained as a
phase of $S_{1 1}$, see Eq.~\eqref{Smatrix}.  For definiteness we
consider $\Gamma_\pi$.  The solution of the single channel equation is
shown in Fig.~\ref{fig:FF}.  The phase $\delta$ approaches $2\pi$ at
large $s$.  Therefore, according to \eqref{asympt}, the solution
decreases as $s^{-2}$.

In the two channel case, there are two independent solutions,
$F_{1 1}(s)$ and $F_{1 2}(s)$. For the
parametrization~\cite{Truong:1989my} the solutions can be constructed
directly as $F_{i j} = T_{i j}/\sqrt{s}$.  The form factor
$\Gamma_\pi$ is a linear combination of type
$C_{11} F_{1 1} + C_{12} F_{1 2}$. At large $s$ we have
$F_{1 1}\sim s^{-1}$, whilst $F_{1 2} \sim s^{-2}$. Therefore we set
$C_{11}=0$ and $C_{12} = m_\pi^2/F_{1 2}(4m_\pi^2)$. The absolute
value of the form factor $\Gamma_\pi$ in the two channel approximation
is shown in Fig.~\ref{fig:FF}.

\begin{figure}[!htb]
\hskip 0.05\columnwidth
\includegraphics[width=0.9\columnwidth]{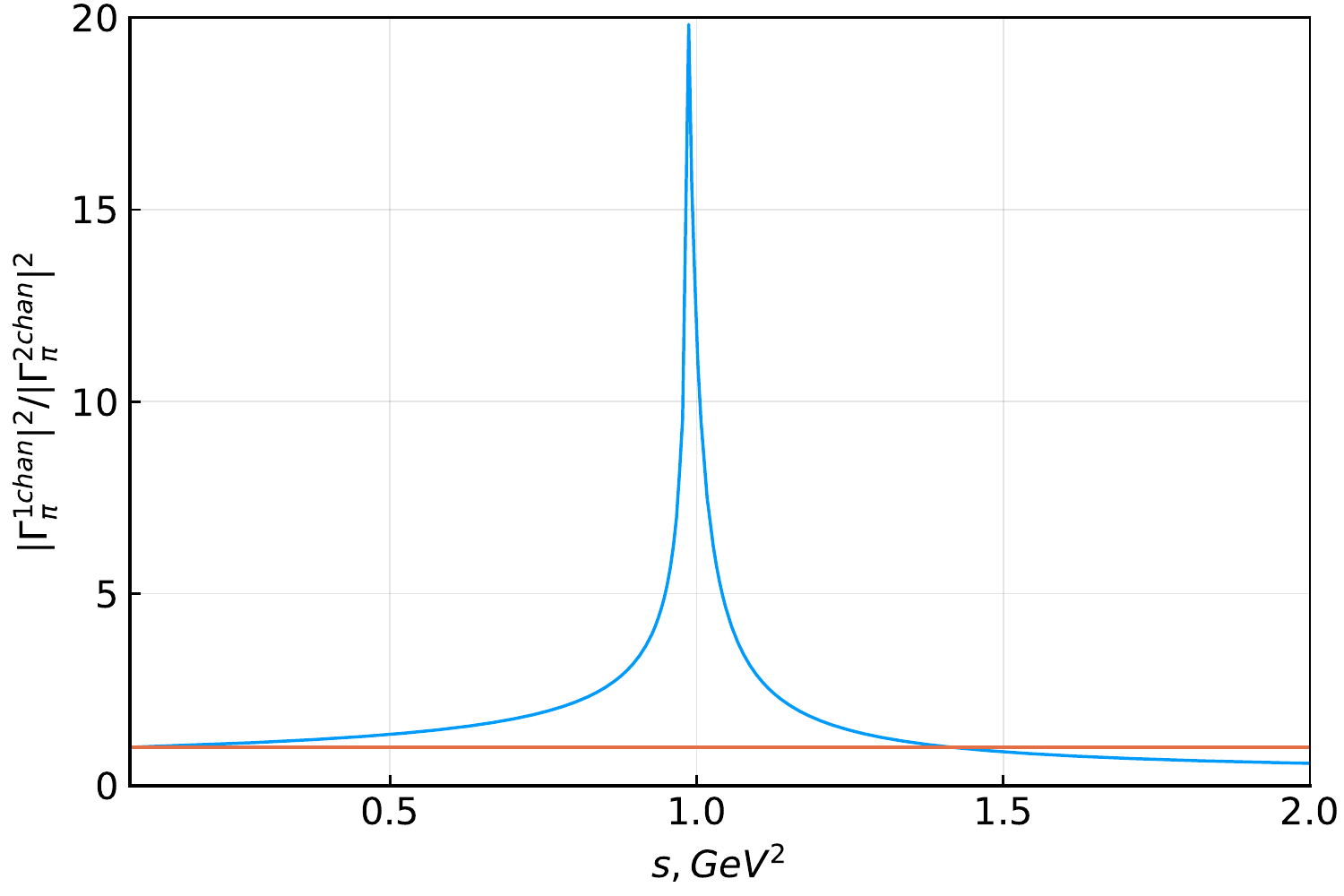}
\caption{The square of the ratio of the absolute values of the form factors as 
a function of $s$.
It is clear that the result is overestimated in the single channel case.
The horizontal line shows the position of $1$.
\label{fig:ratio}}
\end{figure}

As one observes from Fig.~\ref{fig:FF}, the single channel form factor
is much larger around $1$\,GeV. The partial decay width is proportional
to the second power of the form factor, therefore we show the square
of the ratio of the form factors in
Fig.~\ref{fig:ratio}. It is clear that the decay width calculated in
the single channel approximation is overestimated by a factor of
$\simeq 20 $ around $1$\,GeV.

It is important to stress that Figs.~\ref{fig:FF} and \ref{fig:ratio}
are sensitive to the parametrisation of the T-matrix. In fact,
depending on the parametrisation, the blue solid curve in
Fig.~\ref{fig:FF} could be either a peak or even a dip (see Figure 2
in \cite{Ananthanarayan:2004xy}).  The reason of such drastic
uncertainty is the relative position of the $f_0(980)$ resonance and
the inelastic kaon threshold, see an extensive discussion in
\cite{Yndurain:2003vk,Ananthanarayan:2004xy,Yndurain:2005cm,Caprini:2005gd}.
\emph{Nevertheless, our general conclusion is that the naive reduction from
two channels to a single channel does not work.}

The example above demonstrates that the reduction of a $2\times2$
unitary matrix~\eqref{Smatrix} to a $1\times1$ unitary matrix
$\exp(2i\delta)$ leads to significant errors.  Note that in the two
channel case $S_{11} = \eta \exp(2i\delta)$, where the inelasticity
parameter $\eta(s)$ is equal to $1$ below the first
  inelastic
threshold $4m_K^2$ and is less than $1$ above.

\begin{figure}[!htb]
\hskip 0.05\columnwidth
\includegraphics[width=0.9\columnwidth]{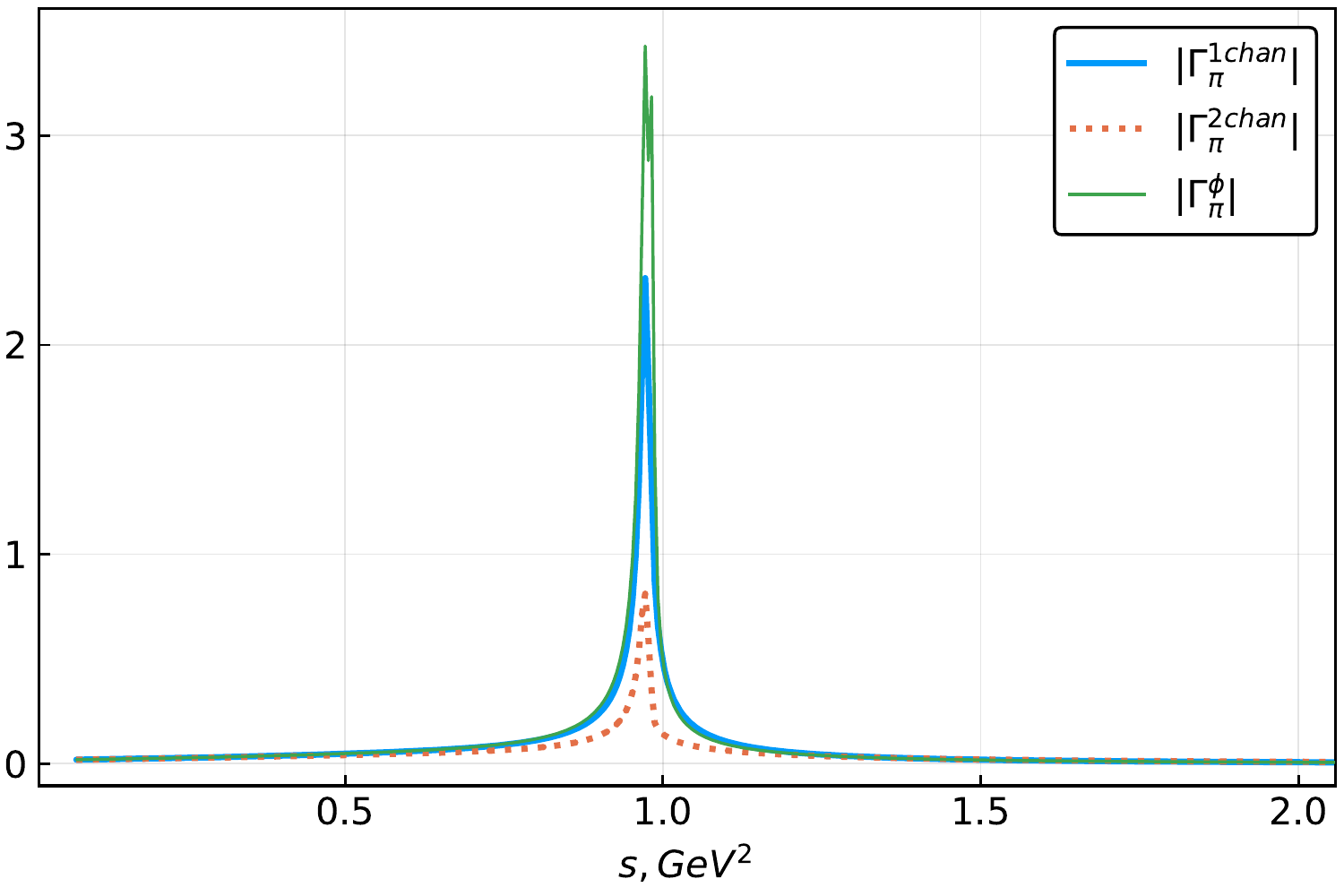}
\caption{Absolute values of the pion form factors as functions of $s$.
The blue solid line represents  $|\Gamma_\pi(s)|$ obtained in the single channel approximation using the phase $\delta$.
The thin green line represents  the same form factor obtained using the phase $\phi$, 
Eq.~\eqref{phi}.
The dotted line is the analogous form factor in the two channel approximation.
\label{fig:FF_phi}}
\end{figure}

In fact, it is possible to account for inelasticity $\eta(s)$ above
the kaon threshold in a better way.  Let us consider
$S_{11} = \eta \exp(2i\delta)$ as a \emph{single channel} scattering
matrix.  It can be rewritten in yet another form
\begin{equation}
  S_{inel}=\eta\, e^{2i \delta}=e^{2 i \left(\delta+i \left[-\frac12 \log \eta\right] \right)}  \equiv e^{2 i \delta_c},
\end{equation}
where we have introduced the complex phase
\begin{equation}
  \delta_c = \delta+i \psi, \quad \psi(s) = \left[-\frac12 \log \eta(s)\right].
\end{equation}
This S-matrix is no longer unitary and it is not legitimate to apply
the arguments based on unitarity in the way it has been done in
Section~\ref{sec:review_of_the_formalism}.  However, it is still
possible (see, e.g.~\cite{barton1965}) to relate the form factor and
the scattering amplitude in the case of a complex phase shift using
the trick of Refs.~\cite{Goldberger:1958vp,Federbush:1958zz}.  This
leads to replacement of the phase shift $\delta$ in the equations with the
effective phase $\phi$ given by~\cite{barton1965}
\begin{equation}
  \tan\phi = \frac{e^{-2 \psi} \sin 2 \delta}{1+e^{-2 \psi} \cos 2 \delta}\,.
  \label{phi}
\end{equation}
The Omn{\`e}s solution~\eqref{Omn_sol} with this phase accounts for
inelasticity.  It is quite remarkable that if a resonance---which
manifests itself as a jump of the phase shift $\delta$ by
$\pi$---occurs in the inelastic region, the phase \eqref{phi} doesn't
change by $\pi$.  This is precisely the reason why the relative
position of the resonance and the kaon threshold is so important.  In
our example the resonance is located before the onset of inelasticity
and the form factor calculated with the phase \eqref{phi} does
not differ very much from the one calculated with $\delta$, see
Fig.~\ref{fig:FF_phi}.  \emph{Therefore, even the improved reduction, which
attempts to imitate heavy channels by inelasticity of the reduced
S-matrix, does not resolve the problem.}

One can argue that in some other parametrisation, e.g.\ the one used
in Ref.~\cite{Ananthanarayan:2004xy}, the procedure of introducing the
complex phase could lead to a resonable agreement between the single
channel and two channel approaches.  This is indeed the case for the
form factor $\Gamma_\pi$. However, this cannot be achieved
simultaneously for all form factors, in particular for $\Delta_\pi$,
which contributes to the decay amplitude in the same way as
$\Gamma_\pi$, see Eq.~\eqref{G}.

\subsection{Scalar radius of the pion} 
\label{sub:scalar_radius_of_the_pion}

At the same time, there are successful predictions that can be
obtained from a reduced 2 channel approach.

The pion form factor $\Gamma_\pi$ has been considered, e.g., in
Refs.~\cite{Moussallam:1999aq,Colangelo:2001df,Yndurain:2003vk,
  Ananthanarayan:2004xy,Yndurain:2005cm,Caprini:2005gd,Oller:2007xd}
in the context of the scalar radius of the pion
$ \langle r^2 \rangle_s^\pi$.
\begin{equation}
  \Gamma_\pi(z) = \Gamma_\pi(0) \left( 1 + \frac16 \langle r^2 \rangle_s^\pi \,z + O(z^2)\right ).
  \label{scalar_radius}
\end{equation}
The scalar radius is related to the $\mathcal{O}(p^4)$ coupling
constant $\bar{\ell}_4$~\cite{Gasser:1983kx} in ChPT and therefore is
of considerable interest.  The result obtained using the dispersive
approach in the two channel approximation \cite{Ananthanarayan:2004xy}
is in agreement with lattice simulations~\cite{Gulpers:2015bba}
and experimental measurements~\cite{Colangelo:2001df}.

The possible explanation of this good agreement, in spite of the
problems described in the previous Section, is the following.  The
scalar radius \eqref{scalar_radius} measures the slope of the form
factor at $s=0$. The form factor can always be represented in the form
\eqref{Omn_sol} where now $\delta$ is the phase of the form factor,
not the scattering phase.  Then, from \eqref{scalar_radius} and
\eqref{Omn_sol} one can get the scalar radius by integrating $\delta$ over $s$.  
The values of $\delta$ at large $s$ are suppressed by $s^2$, so the scalar
radius is rather insensitive to the dynamics at high energies. 
Nevertheless, there was an active discussion in the
literature~\cite{Ananthanarayan:2000ht,Yndurain:2003vk,
  Ananthanarayan:2004xy,Yndurain:2005cm,Oller:2007xd} about the
importance of the two channel approach.



\section{Discussion}
\label{sec:Conclusion}
In this work we have investigated the impact of the final state
interaction on the width of a light Higgs-like scalar.  The decay
width to mesons is proportional~\cite{Donoghue:1990xh} to
$|G_i(s)|^2$, where $G_i(s)$ is the linear combination~\eqref{G} of
three form factors defined in Eq.~\eqref{FF_definition} and
$i = \pi, K, \ldots$.  Most of the analyses\footnote{The only
  exception is the work by Moussallam~\cite{Moussallam:1999aq}, where
  four pion final state has been effectively accounted for using the
  parametrisation of Refs.~\cite{Kaminski:1997gc,Kaminski:1998ns}.
  However, he found that the available experimental data are not
  enough to fix the three channel parametrization, and different choises of the 
  model parameters lead to significantly different results.}  have been
limited to the two channel case, namely, $i = \pi, K$.  Though all
data relevant for the two channels (two scattering phases and
inelasticity) have been encoded in a number of phenomenological
parametrizations
\cite{Au:1986vs,Kaminski:1997gc,Kaminski:1998ns,Hoferichter:2012wf},
we argue that there is an intrinsic uncertainty related to neglected
channels.  This can be most easily inferred from
Eqs.~\eqref{Reduced_system}. There is a contribution from neglected
channels which can be understood as an exchange of heavier states in
the loops and which is simply dropped in the two channel analysis.
This contribution is suppressed for quantities defined at small $s$,
such as scalar radius of the pion, but becomes more and more important
once one goes away from $s=0$. The decay width of a light scalar boson
$S$ to hadrons as a function of $m_S$ is extremely sensitive to the
effects of higher channels.

There is yet another source of uncertainty in the application of
dispersion approach to the calculation of the form factors. This
uncertainty stems from the fact that the off-diagonal elements of the
T-matrix are needed in the \emph{unphysical} region.  As we have
discussed in section \ref{sec:review_of_the_formalism}, $T(s)$ cannot
be directly extracted from the experimental data and therefore is
defined as an approximation, which depends on a particular analytic
parametrization of the S-matrix.

To summarize, the method of Ref.~\cite{Donoghue:1990xh} should be
applied with great caution. Moreover, as we have demonstrated, the
present form of the 
reduction of the number of channels leads to significant uncertainties
which are beyond theoretical control. We believe that this fact should
be clarified when exclusion plots or projected sensitivities for the
hypothetical light scalars are presented.
\newline

The work of  I.T. was  partially supported
 by the \mbox{ERC-AdG-2015 grant 694896}.

\bibliography{refsMO}

\end{document}